\documentstyle[aps,draft,prl,twocolumn]{revtex}
\begin{document}
\widetext
\title{Grazing Incidence Infrared Reflectivity of  
La$_{1.85}$Sr$_{0.15}$CuO$_4$ and NbN.}

\author{H. S. Somal,$^1$ B. J. Feenstra,
$^1$ J. Sch\"utzmann,$^1$ Jae Hoon Kim,$^{1,2}$ 
Z. H. Barber,$^3$ V. H. M. Duijn,$^4$ N. T. Hien,$^4$, A. A. Menovsky,$^4$ 
Mario Palumbo, $^5$ and D. van der Marel$^1$}

\address{$^1$ Materials Science Centre,Laboratory of Solid State Physics,
University of Groningen,\\
Nijenborgh 4, 9747 AG Groningen, The Netherlands.\\
$^2$ Dept. of Physics, Yonsei University, Seoul 120-749, Korea.\\
$^3$Dept. of Materials Science and Metallurgy, University of Cambridge, United Kingdom.\\
$^4$Van der Waals-Zeeman Laboratory, University of Amsterdam, The Netherlands.\\
$^5$ Physikalisches Institut, University of Bayreuth, 
D-95440 Bayreuth,Germany. }

\date{15 september 1995}

\maketitle

\begin{abstract}
Infrared reflectivity measurements, using p-polarized light at a
grazing angle of incidence, show an increased
sensitivity to the optical conductivity of highly reflecting
superconducting materials. We demonstrate that when this measurement
technique is applied to the conventional s-wave superconductor NbN,
the results are in perfect agreement with BCS theory.
For the in-plane response of a
La$_{1.85}$Sr$_{0.15}$CuO$_4$ single crystal, 
in the superconducting state, we find a reduction of
the optical conductivity in the frequency range below 20 meV.
The observed frequency dependence 
excludes an isotropic s-wave gap, but agrees
well with model calculations assuming a d-wave order parameter.
\end{abstract}


The symmetry and magnitude of the superconducting order 
parameter belong to the key elements of theoretical models of
the superconducting state of the high T$_c$ cuprates. The
issue of the order parameter-symmetry is intimately related to
the question as to whether superconductivity is caused by
local exchange correlations, anti-ferromagnetic 
spin-fluctuatons, inter-layer pairing or
electron-phonon interactions.\cite{theories}
Much of the data which have been interpreted as evidence for
d-wave pairing, using a wide variety of experimental 
techniques, were obtained on the double-layer compounds
Bi$_2$Sr$_2$Ca$_1$Cu$_2$O$_8$ and YBa$_2$Cu$_3$O$_7$.\cite{Wollman} 
However, for compounds with
two or more CuO layers per unit cell the presence of several
coupled charge reservoirs
complicates the picture, and increases the number of
possible interpretations of the experimental results \cite{Kresin}, 
as was recently discussed\cite{Golubov} for {\em e.g.} the experimentally 
observed sign reversals of the order 
parameter in YBa$_2$Cu$_3$O$_7$ . On the other hand the single
layer cuprate superconductors, 
such as La$_{1-x}$Sr$_x$CuO$_4$, have only a
single band of electrons per unit cell, which in principle
facilititates the interpretation of experimental data.

The experimental results on La$_{1-x}$Sr$_x$CuO$_4$ 
using Raman spectroscopy \cite{Chen},
specific heat measurements\cite{Momono} and inelastic neutron
scattering\cite{birgeneau} indicate a strong 
anisotropy of the order parameter, possibly due to d-wave
pairing. In this Letter we discuss a set of 
infrared experiments on La$_{1.85}$Sr$_{0.15}$CuO$_4$. 
The classical technique of infrared 
spectroscopy offers the advantage of sensitivity to the
bulk of the material, 
high energy resolution and the fact that the optical conductivity 
determined by this method can be interpreted in a theoretically
well-defined way as the current-current correlation function. 

Although mm-wave studies of the low-frequency penetration depth 
have been successfully used to probe, via the temperature dependence,
the possible existence of nodes in the 
order parameter of YBa$_2$Cu$_3$O$_7$, a direct
spectroscopic determination of the order parameter with infrared/mm-wave
techniques has been
limited so far by several experimental factors: (1) The low frequency 
limiting behavior of the reflectivity of a (non-superconducting) metal 
is given by the Hagen-Rubens expression 
$R=1 - 0.366$ ($\nu \rho_{DC})^{1/2}$ with $\nu$ in units of cm$^{-1}$,
$\rho$ in units of $\Omega cm$, the reflectivity of a
good conductor is high. For the high T$_c$ cuprates $\rho\approx
100 (\mu\Omega cm)$, so that for frequencies of the order of $3.5 k_B T_c$,
and $T_c=30 K$, R=0.97. As a result the changes in reflectivity upon
entering the superconducting state are difficult to detect, especially
because the changes take place gradually \cite {Kamaras}.
(2) The range of order parameter
induced changes in conductivity coincides with the frequency range
of optical phonons. The electronic conductivity in the c-direction
is very small in these materials, therefore changes are masked by the rather
strong contributions due to optical phonons \cite{Tamasaku,Kim}. \\
In this paper we describe and demonstrate that the situation with respect
to problem (1) can be improved by using polarized light incident at a
grazing angle. For p-polarised light the reflectivity is given by
the expression:
\begin{equation}
R_p(\theta,\omega)=
           \left|\frac{\epsilon\cos\theta-\sqrt{\epsilon-\sin^2\theta}}
                      {\epsilon\cos\theta+\sqrt{\epsilon-\sin^2\theta}} 
           \right|^2
\end{equation}
For a metal in the limit where $\omega\tau \ll 1$ the absorptivity 
$A_p=1-R_p=\frac{4}{\cos\theta}\sqrt{\frac{\omega}{4\pi\sigma}}$ is
enhanced by a factor $1/\cos\theta$, up to an angle 
$\theta_c=\arccos\sqrt{\frac{\omega}{4\pi\sigma}}$ where it
reaches a maximum . On the other hand,
in the superconducting state the dielectric properties
at low frequencies are dominated by the real part of the
dielectric function, 
$\mbox{Re}(\epsilon)=-c^2(\omega\lambda(T))^{-2}$,
where $\lambda(T)$ is the penetration depth. 
The absorbtivity $A_p=\frac{4}{\cos\theta}
c^{-3}\omega^2\lambda(T)^3 4\pi\sigma$ vanishes as
$\sigma \rightarrow 0$ below the gap.
Therefore, by measuring the reflectivity 
at a grazing angle of incidence with p-polarized light
we enhance our sensitivity to changes in $\sigma$ with roughly a factor
of $1/cos\theta$. In the present study 
we have chosen an angle $\theta=80\pm3^o$,
resulting in an enhancement factor of 
approximately 6. For the samples studied in this
paper the superconductivity induced changes in reflectivity are of the
order of 1 \% at normal incidence. This is at the border of what can be
measured experimentally (with a typical accuracy of about $0.5\%$). At 
$\theta$ =$80 ^o$ the level of superconducting induced changes is raised 
to a comfortable $10 \%$. The grazing incidence method is 
different from and complementary to improvements in counting 
statistics, {\em e.g.} by using larger samples, brighter sources, etc. 

La$_{1.85}$Sr$_{0.15}$CuO$_4$ single crystals with a T$_c$ of 32 K
were grown by the traveling solvent floating zone method,
and cut along the a-c plane. The dimensions of the clean and polished crystal 
surface were approximately 7 mm along the 
a-axis and 4 mm along the c-axis. The samples
were mounted such, that the plane of scattering coincides with the
CuO-planes. The samples were mounted in a conventional
flow cryostat using the two opposite windows for scattering in a cone
of $80\pm3^o$ relative to the surface normal. A wire grid polarizer was
used to select the proper polarization. Special care was taken to
prevent stray light from reaching the detectors. Although all the
measurements reported here were taken with p-polarized light, by
rotating the wire-grid polarizer by $90^o$, we were able to
direct $\vec{E}\parallel\vec{c}$ and detect the appearance of the
c-axis plasmon as $T$ drops below $T_c$ \cite{Kim}. This was used both as an 
{\em in situ} check for correct crystal orientation and as an 'internal'
indicator of the sample temperature. 
NbN films of 400nm and 55nm thickness were prepared 
on substrates of MgO by DC reactive magnetron sputtering,
in a gas composition of 
3.0\% CH$_4$ / 30.0 \% N$_2$ / Ar,  at a temperature of
approximately
840 $^\circ$ C to  produce high quality epitaxial growth \cite{Barber}.
The  carbon addition improves film quality
which was checked by X-ray diffraction and high resolution 
electron microscopy. 
The films had a superconducting transition 
temperature of 16.5 K, and a resistivity ratio 
(300 K / 16.5K) close to unity.

Reflectivities at 80 degrees angle of incidence were measured at
temperatures between 4 and 60 K. With a gold reference mirror we
checked that the effects of thermal expansion of the cryostat
on our results is very small within this range of temperatures. 
Initially the normal incidence absolute reflectivity at 40 K was 
measured with  high precision 
(inset of Fig. 1, upper panel), we then calculated
the dielectric function using the  Kramers-Kronig transformation,
and from this the reference reflectivity curve at 
40 K and 80 degrees angle of incidence. 
Absolute reflectivities at all temperatures
were obtained by multiplying the reflectivity ratios to this
reference curve, and by correcting for thermal expansion 
effects of the cryostat as detected with a gold reference mirror
in a separate session. Although under conditions of s-polarization
the c-axis plasmon shows up very prominently below T$_c$, with
$\vec{E}$ polarized parallel to the planes we observed no distinct
gap feature in the case of 
La$_{1.85}$Sr$_{0.15}$CuO$_4$. Because the observed 
thermally induced changes are quite subtle, 
we reproduced the measurements several times on two different crystals.

The absolute reflectivities are displayed in the upper panel of Fig. 1 for 
temperatures between 4 and 50 K. In the inset we show the reflectivity taken
at normal incidence. 
Below 3 meV the extrapolation was based upon a two fluid model
as described in \cite{Marel}. A Kramers-Kronig transformation gives the
pseudo-dielectric function at 80$^o$, and from this the dielectric
function and the optical conductivity which is displayed in 
the lower panel of Fig. 1.
In order to check the internal consistency of the analysis, as well
as agreement with published values of the penetration depth, we
calculated $\lambda(\omega)$ by two different methods. The skin
depth, defined as
$\delta=c/(\omega\mbox{Im}\sqrt{\epsilon})$, must be taken 
in the limit $\omega\rightarrow 0$ to obtain the superconducting
penetration depth (inset of Fig. 1, lower panel). 
Alternatively, if the temperature dependence results from a transfer
of spectral weight to the condensate peak at $\omega=0$, the
penetration depth follows from the Glover-Tinkham-Ferell
sumrule formula $\lambda(\omega)^{-2}=
8\int_{0^+}^\infty(\sigma_n(\omega')-\sigma_s(\omega'))d\omega'$.
In our analysis these two values coincide with an accuracy better than 1 \%. 
We therefore conclude, that the 'missing' area in the conductivity is
indeed transferred to the condensate peak, and is therefore
inherently linked to the formation of the superconducting state.
The values of $\lambda=4300 \pm 30\AA$  and
$4900 \pm 30\AA$ at 4 and 20 K respectively are in good
agreement with measurements at 10 GHz\cite{Shibauchi}. 

It has been suggested\cite{Gao}, that the low
frequency electrodynamics can be understood as a sum of a
temperature independent mid-infrared band and
a Drude contribution which narrows upon lowering the temperature.
This empirical decomposition of the data
has been used to describe normal
incidence reflectance and conductivity spectra for frequencies
typically above  12 meV. 
For our sample the mid-infrared band can be parameterized with the function
$\sigma_{M}\mbox{Re}\omega/(\omega+i[\omega_{M}^2-\omega^2]\tau)$ where
$\sigma_{M}=800$ (S/cm), $\hbar\omega_{M} \approx$ 0.16 (eV)
and $\hbar\tau^{-1}=$ 0.66 (eV). Clearly for $\hbar\omega < 40$ (meV) 
the conductivity is dominated
by the free carrier contribution. The suppression of conductivity 
below 18 meV is markedly different from a narrowing (or depletion)
of the Drude peak on a constant MIR background. On the other hand
there is also no gap in the observed frequency 
range. If La$_{1.85}$Sr$_{0.15}$CuO$_4$
would be an isotropic s-wave superconductor, 
then BCS theory predicts that the 
gap should be at  approximately 10 meV. As there is no spurious background
in the present spectra, it is clear from these
data that such a BCS gap is absent. 

To verify, that our experimental method will indeed
show the presence of a gap if there is one, we show in Fig. 2 the
conductivity obtained with the same grazing incidence technique
on a thick (400 nm) film of NbN.
NbN thin films deposited on MgO are suitable for IR transmission
spectroscopy, our measurements show that this material behaves
like a classical BCS superconductor, as can be seen from the quality
of the fits in Fig. 2 (lower panel).
With $T_c=16.5K$
the gap is at approximately 6 meV (2$\Delta/ k_BT_c \simeq$ 
4, which is
now quite close to the lower limit of our window of observation
(of 3 meV). In terms of 'cleanliness' in the normal
state, this material has an even higher reflectivity than high T$_c$
superconductors as the DC conductivity is about 14000 
$(S/cm)$. In this case, however, the changes in
the measured reflectivity at grazing 
incidence are quite large, and a clear gap opens up
in the conductivity. As can be seen from the solid curves
the agreement with  simulations \cite{Zimmerman} assuming an isotropic 
s-wave gap of $2\Delta=4k_BT_c$ for NbN \cite{Karecki}
with intermediate scattering rate ($\hbar/\tau=14$ meV))
is excellent. 

In Fig. 3 we display our experimental data at 4 K and 40 K together
with model calculations 
for La$_{1.85}$Sr$_{0.15}$CuO$_4$, based on
BCS s-wave theory with finite scattering ($2\Delta=3.5 k_BT_c$)
\cite{Zimmerman} and 
a weak coupling calculation with finite scattering rate assuming
d-wave pairing \cite{Graf}. As parameters we took $\hbar/\tau=$ 6 meV,
and $\omega_p =$ 0.73 eV. By fixing T$_c$
at 32 K we also fixed the value of the gap. For the
case of d-wave pairing T$_c$ is quite strongly
suppressed due to the presence of scattering\cite{radtke}. 
In our calculations
we had to take $T_{c0}=64 $K, which is then reduced to 32 K
due to our assumed scattering rate of 6 meV. There is an
intrinsic ambiguity in the choice of the latter: Although the
scattering rate needed to fit the data for $T<50$ K and
$\omega > $ 6 meV is temperature independent, the
linear temperature dependence of the decay rate does still
exist at lower (notably DC) frequencies. Hence the physical
origin of the scattering rate seems not to be solely due to 
impurity
scattering. The present weak-coupling calculations do not
faithfully represent this aspect of the problem, but have
the advantage of using only a limited number
of adjustable parameters. As $T_c$, $1/\tau$ and $\omega_p$
are fixed by experimental constraints, the only free parameter
left is the scattering cross section  $\sigma_{sc}=\sin^2{\eta}$,
where $\eta$ is the phase shift. 
The comparison with experimental data in Fig. 3 strongly
favours d-wave symmetry with $\sigma_{sc}=0.4$, so that
$\eta\approx 0.2\pi$. The analysis of microwave surface impedence measurements
of YBa$_2$Cu$_3$O$_y$ between 4 GHz - 40 GHz indicates
that for these materials the elastic scattering is very small, and
close to the unitarity limit ($0.98 < \eta < 1$) \cite{hirschfeld,jacobs}.
The temperature dependence of the ab-plane penetration depth 
of La$_{2-x}$Sr$_x$CuO measured at 10 GHz
drops slightly below the BCS-prediction for s-wave
pairing \cite{Shibauchi}. These data do not exclude the possibility
of d-wave pairing in the presence of impurity scattering,
in which case $\Delta\lambda \propto T^2$\cite{goldenfeld}. 
Although an isotropic s-wave
gap is clearly ruled out by our analysis, we can not rule
out the possibility of a strongly anisotropic gap, or an
incoherent mixture of s and d-wave symmetry ('s+id'). In such
cases a lower bound exists on the distribution of gap values
along the Fermi-surface. In the optical spectra a full gap
only opens below the lower bound of this distribution. From
our data we conclude, that such a lower bound, if it
exists, must be located below 3 meV (1.3 k$_B$T$_c$). 

Using grazing incidence reflectivity, we determined the optical
conductivity of La$_{1.85}$Sr$_{0.15}$CuO$_4$ 
and NbN below and near the superconducting
phase transition down to a photon energy of 3 meV. In the
case of NbN a clear BCS gap was observed. 
For La$_{1.85}$Sr$_{0.15}$CuO$_4$ good
agreement was obtained with model calculations assuming
d-wave pairing and scattering closer to the Born-limit.

{\em Acknowledgements}
We gratefully acknowledge D. Rainer and M. Graf for their
theoretical support. This investigation 
was supported by the Netherlands Foundation for
Fundamental Research on Matter (FOM) with financial aid from
the Nederlandse Organisatie voor Wetenschappelijk Onderzoek (NWO).

\begin{figure}
\caption{Upper panel: Reflectivity of La$_{1.85}$Sr$_{0.15}$CuO$_4$ 
at 80 degrees angle of
incidence at 4 K (solid), 20 K (dashed), 27 K (dotted), 
40 K(dash-double-dotted) and 50 K (chained). 
Inset: Reflectivity at normal
angle of incidence for 40 K(solid) and 300 K (dashed). 
Lower panel: Optical conductivity at 50 K (closed triangles), 
40 K (circles), 2 K (lozenges), 20 K (closed triangles) and 4 K (squares)  
obtained from the reflectivities using Kramers-Kronig 
relations. Inset: Skin depth for 4 (solid) and 20 K (dashed). The
solid dots are the penetration depth obtained from the f-sum rule. }
\label{fig1}
\end{figure}
\begin{figure}
\caption{Upper panel: Optical conductivity of NbN calculated from 80 degree
angle of incidence reflection spectra for 9 (closed circles), 13 (open squares)
and 18 K (closed diamonds). Solid curves:
Theoretical calculations assuming isotropic s-wave pairing. Lower panel:
Measured transmittivity ratios of a 55 nm thick NbN film on MgO
(closed circles: T(9 K)/T(18 K), open symbols: T(13 K)/T(18 K)) 
The solid curves are calculated using
the same parameters as for the conductivity of the top panel.}
\label{fig2}
\end{figure}
\begin{figure}
\caption{Comparison of the experimental optical conductivity at 4 K
(solid circles) and 40 K (open circles) to
model calculations for the optical conductivity 
of La$_{1.85}$Sr$_{0.15}$CuO$_4$ assuming an 
intermediate impurity scattering rate at
T=40 K (dash-double-dotted). The other curves are calculated for 4 K
assuming isotropic s-wave pairing (chained), 
d-wave pairing with $\sigma=0.1$ (dotted),
d-wave with $\sigma=0.4$ (solid), d-wave with $\sigma=0.8$ (dashed).}
\label{fig3}
\end{figure}
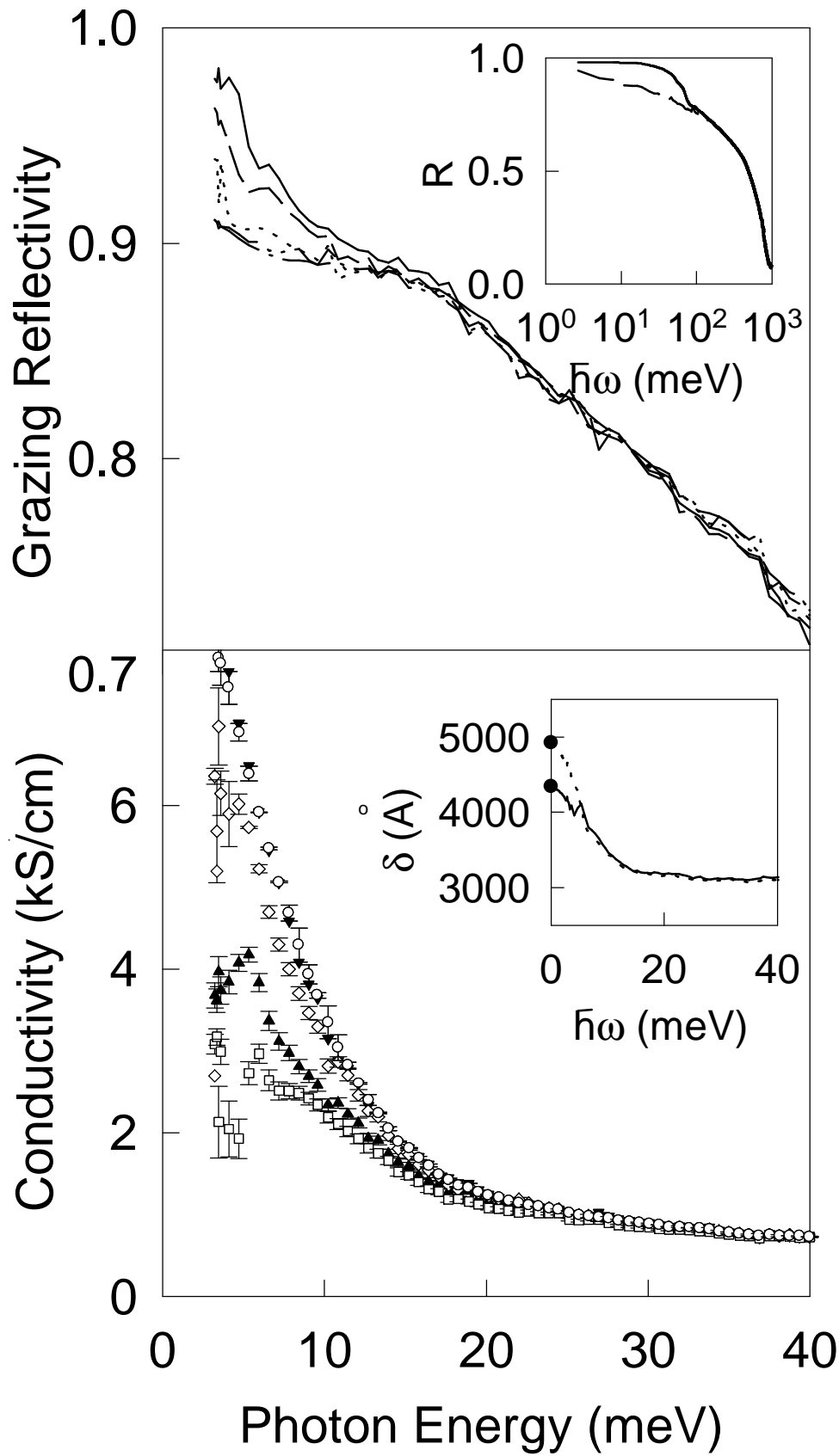
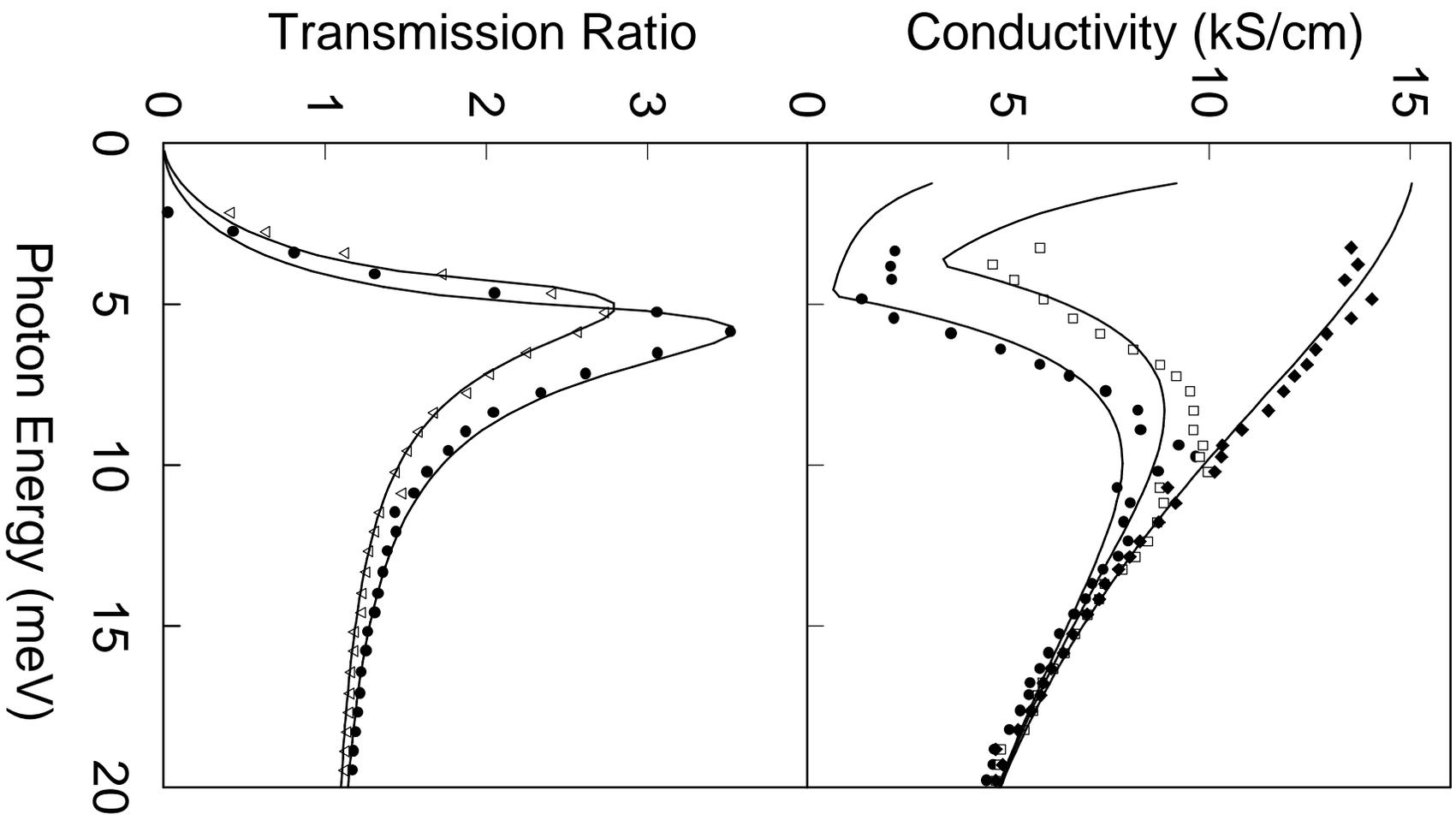
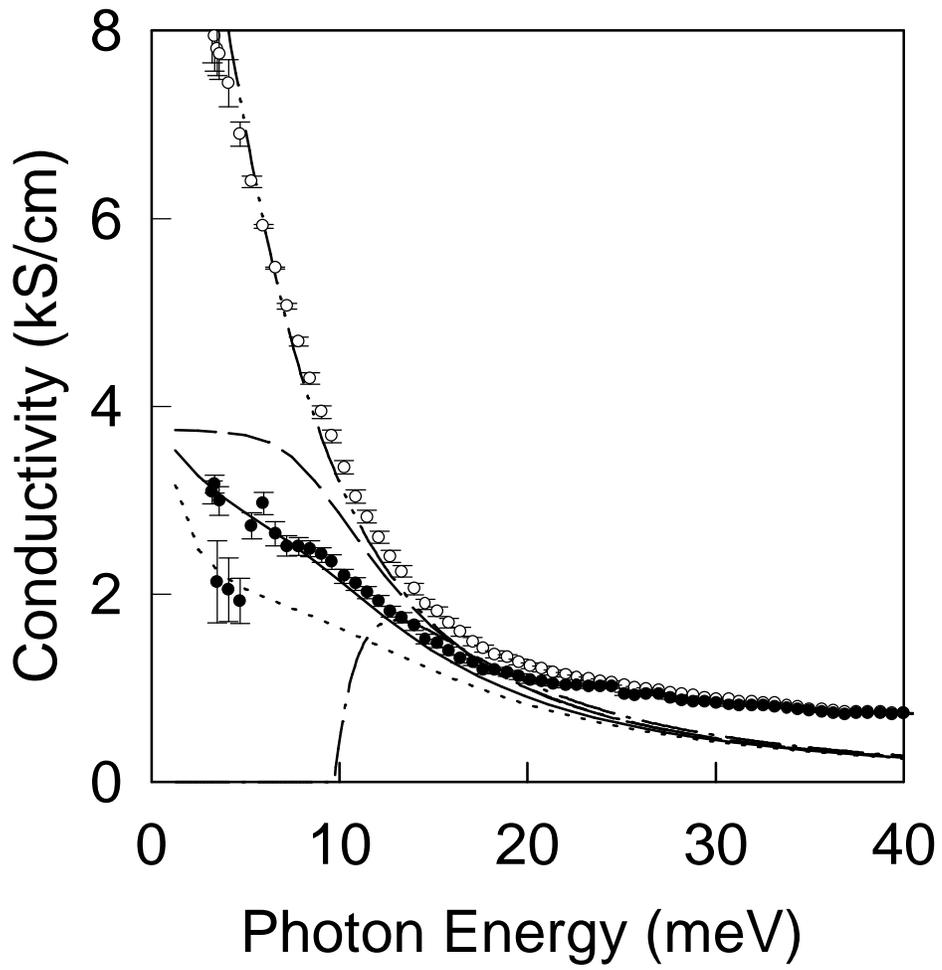
\end{document}